\RequirePackage{etoolbox}
\csdef{input@path}{%
 {sty/}
 {img/}
}%
\csgdef{bibdir}{bib/}

\documentclass[ba]{imsart}
\pubyear{0000}
\volume{00}
\issue{0}
\doi{0000}
\firstpage{1}
\lastpage{1}

\usepackage{amsthm}
\usepackage{amsmath}
\usepackage{natbib}
\usepackage[colorlinks,citecolor=blue,urlcolor=blue,filecolor=blue,backref=page]{hyperref}
\usepackage{graphicx}
\usepackage{hhline}
\usepackage{multirow}
\usepackage{subcaption}

\startlocaldefs
\numberwithin{equation}{section}
\theoremstyle{plain}

\endlocaldefs

\begin{document}


\begin{frontmatter}
\title{Frequency Coverage Properties of a Uniform Shrinkage Prior Distribution}

\runtitle{Frequency Coverage Properties of a Uniform Shrinkage Prior}
\begin{aug}
\author{\fnms{Hyungsuk} \snm{Tak}\thanksref{addr1}\ead[label=e1]{hyungsuk.tak@gmail.com}}

\runauthor{Tak and Morris}

\address[addr1]{
     Statistical and Applied Mathematical Sciences Institute, \printead{e1} 
}

\end{aug}

\begin{abstract}
A uniform shrinkage prior (USP) distribution on the unknown  variance component of a random-effects model is known to produce good frequency properties.  The USP has a parameter that determines the shape of its density function, but it has been neglected whether the USP can maintain  such good frequency properties regardless of the choice for the shape parameter. We investigate which choice for the shape parameter of the USP produces Bayesian interval estimates of random effects that  meet their nominal confidence levels better than several existent choices in the literature. Using univariate and multivariate Gaussian hierarchical models, we empirically show that  the USP can achieve its best frequency properties when its shape parameter makes the USP behave similarly to an improper flat prior distribution on the unknown variance component.
\end{abstract}


\end{frontmatter}



\section{Introduction}\label{sec:intro}
We briefly review a uniform shrinkage prior (USP) using a multivariate Gaussian hierarchical model (or a  multivariate Normal-Normal model) \citep{daniels1999prior, everson2000inference}. We use notation $\boldsymbol{Y}\!_j=(Y_{1j}, Y_{2j}, \ldots, Y_{pj})^\top$ to denote a vector of $p$ unbiased estimators of   group~$j$ for a vector of $p$ random effects~$\boldsymbol{\theta}_j=(\theta_{1j}, \theta_{2j}, \ldots, \theta_{pj})^\top$ with  a two-level structure as follows. For $j=1, 2, \ldots, k$,
\begin{align}
\boldsymbol{Y}\!_j\mid \boldsymbol{\theta}_j & \stackrel{\textrm{indep.}}{\sim} \textrm{N}_p(\boldsymbol{\theta}_j,~ \boldsymbol{V}\!_j),\label{twolevel_structure1}\\
\boldsymbol{\theta}_j\mid \boldsymbol{A}, \boldsymbol{\beta} &\stackrel{\textrm{indep.}}{\sim}  \textrm{N}_p(\boldsymbol{X}\!_j^\top \boldsymbol{\beta},~ \boldsymbol{A}),\label{twolevel_structure2}
\end{align}
where $\boldsymbol{V}\!_j$ is a known $p\times p$ covariance matrix, $\boldsymbol{\beta}=(\beta_1, \beta_2, \ldots, \beta_{mp})^\top$ is a vector for  unknown $mp$ regression coefficients, $\boldsymbol{X}\!_j$ is a known $mp\times p$ block diagonal matrix with $\boldsymbol{x}_j=(x_{j1}, x_{j2}, \ldots, x_{jm})^\top$, a  covariate vector of length $m$ for group $j$, repeatedly appearing along the diagonal, and $\boldsymbol{A}$ is an unknown $p\times p$ covariance matrix of the conjugate Gaussian prior distribution. The corresponding conditional posterior distribution for group $j$ is 
\begin{equation}\label{multivariate_conditional_posterior}
 \boldsymbol{\theta}_j\mid \boldsymbol{A}, \boldsymbol{\beta}, \boldsymbol{y}_j\sim \textrm{N}_p(~(I_p-\boldsymbol{B}_j)\boldsymbol{y}_j+\boldsymbol{B}_j\boldsymbol{X}\!_j^\top \boldsymbol{\beta},~(I_p-\boldsymbol{B}_j)\boldsymbol{V}\!_j~),
\end{equation}
where $\boldsymbol{y}_j$ is the observed unbiased estimate, $I_p$ is a $p$-dimensional identity matrix, and $\boldsymbol{B}_j\equiv \boldsymbol{V}\!_j(\boldsymbol{V}\!_j+\boldsymbol{A})^{-1}$ is a positive definite $p\times p$ matrix for the shrinkage factor of group $j$.  \cite{daniels1999prior} defines the multivariate USP as a Uniform distribution on the space of the positive definite shrinkage matrix $\boldsymbol{B}_0=\boldsymbol{V}\!_0(\boldsymbol{V}\!_0+\boldsymbol{A})^{-1}$ whose eigenvalues are between $(0, 1]$; $\boldsymbol{V}\!_0$ is a known positive definite symmetric matrix.  The density function of the USP with respect to  $\boldsymbol{A}$ that is transformed from the USP on $\boldsymbol{B}_0$  is proportional to
\begin{equation}\label{usp}
\vert \boldsymbol{V}\!_0 + \boldsymbol{A}\vert ^{-p-1}\times I_{\{\vert \boldsymbol{A}\vert >0\}}d\boldsymbol{A}, 
\end{equation} 
where $I_{\{w\}}$ is an indicator function of $w$. The known matrix $\boldsymbol{V}\!_0$ determines the shape of this density function in \eqref{usp} and thus we call it the shape parameter of the USP. (Hereafter, a USP is always with respect to $\boldsymbol{A}$.) An approximate USP in generalized linear mixed models \citep{natarajan2000reference},  derived by replacing the first-level distributions for $\boldsymbol{Y}\!_j$  with an MLE-based Gaussian approximation, is the same as \eqref{usp} with a specific value for $\boldsymbol{V}\!_0$. (See also \cite{chen2016approximate} for the approximate USP in multivariate generalized linear mixed models.) For a univariate case ($p=1$), we use unbolded notation, i.e., $Y_j$, $y_j$, $\theta_j$, $V_j$, $V_0$, $A$,  $B_j$, and $B_0$.

The USP is known to have several attractive properties.  \cite{strawderman1971proper} showed using a Normal-Normal model ($p=1$) that the unconditional posterior mean of a random effect, i.e., $E(\theta_j\mid y_j)$, based on the USP leads to an admissible minimax estimator if $k>5$. Constructing a  Poisson-Gamma hierarchical model ($p=1$), \cite{christiansen1997hierarchical} indicated that the USP is relatively non-informative with good frequency coverage properties in random effects estimation. In particular, they mentioned that the shape parameter of the USP could be adjusted to have more conservative results in estimating random effects, which we empiricially prove in this article. Using a  Normal-Normal model ($p=1$), \cite{daniels1999prior} compared the frequency properties of a USP with those of several non-informative priors such as improper flat and Jeffreys' priors  under the settings for hypothesis testing and estimation on $A$. \cite{natarajan2000reference} showed good frequency coverage properties of the approximate USP compared to other default priors, e.g., approximate Jeffreys' prior, using  non-Gaussian data.

Many authors have used various choices for the shape parameter  of the USP. \cite{albert1988computational} set $V_0=1$  for a hierarchical generalized linear model, although there was no specific reason for this choice; the article focused on an accurate approximation to posterior distributions of $A$ and $\boldsymbol{\beta}$.  \cite{christiansen1997hierarchical} suggested setting the shape parameter to a median value of the random-effects variance component for their Poisson-Gamma model as a less informative choice. \cite{dumouchel1994hierarchical} suggested setting $V_0$  to a harmonic mean of $\{V_1, V_2, \ldots, V_k\}$  as a typical value of $V_j$ in a univariate Normal-Normal model. Following \cite{dumouchel1994hierarchical}, \cite{daniels1999prior}  set $\boldsymbol{V}\!_0$ to the harmonic mean of $\{\boldsymbol{V}\!_1, \boldsymbol{V}\!_2, \ldots, \boldsymbol{V}\!_k\}$  in a multivariate Normal-Normal model. This harmonic mean was also adopted for the approximate USP \citep{natarajan2000reference}. Besides the harmonic mean, \cite{everson2000inference} set $\boldsymbol{V}\!_0$ to an arithmetic mean  of $\{\boldsymbol{V}\!_1, \boldsymbol{V}\!_2, \ldots, \boldsymbol{V}\!_k\}$ as another typical value of $\boldsymbol{V}\!_j$  to mimic an equal variance case. 

\begin{figure}[b!]
\begin{center}
\includegraphics[width = 200pt, height = 120pt]{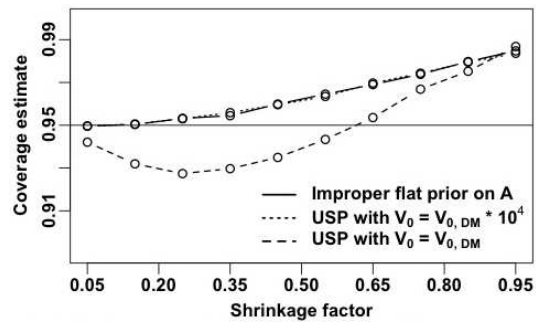}
\caption[]{A partial result of our numerical illustration in Section~\ref{sec4school} shows that there exists a  choice for the shape parameter of the USP that meets the spirit of 95\% confidence level better than an existent choice in the literature, e.g., $V_{0, \textrm{DM}}$ denotes a harmonic mean of $\{V_1, V_2, \ldots, V_k\}$ suggested by \cite{dumouchel1994hierarchical}. See Section~\ref{sec4school} for details.} 
\label{pre_res}
\end{center}
\end{figure}

However, it has been neglected whether the USP maintains   good frequency properties regardless of the choice for its shape parameter, and no one has investigated the frequency coverage properties of the USP in a multivariate case ($p\ge2$). This article investigates the frequency coverage properties of the USP according to the shape parameter, using univariate and bivariate Normal-Normal models. To do this, we extend a univariate repeated sampling coverage evaluation \citep{tak2016b} to a multivariate one to see which choice for the shape parameter produces Bayesian interval estimates for random effects that meet their nominal confidence levels better.  Our numerical illustrations show that the best frequency coverage properties of the USP can be achieved when the shape parameter lets the USP  approach an improper flat prior distribution on~$\boldsymbol{A}$. Figure~\ref{pre_res} displays one numerical illustration in Section~\ref{sec4school}. It shows that a Normal-Normal model based on a USP with $V_0$ set to a harmonic mean of $\{V_1, V_2, \ldots, V_k\}$ denoted by $V_{0, \textrm{DM}}$ \citep{dumouchel1994hierarchical, daniels1999prior, natarajan2000reference} produces posterior intervals for random effects that  do not meet the 95\% confidence level. On the other hand, a choice for $V_0$ that flattens the USP on $A$ with $V_0=10^4\times V_{0, \textrm{DM}}$ achieves the spirit of the 95\% confidence level better. See Section~\ref{sec4school} for details.



The rest of this article is organized as follows. We introduce a multivariate version of the frequency coverage evaluation in Section \ref{sec3}.  In Section \ref{sec4}, we explain which choices for the shape parameter achieve the spirit of the confidence level better and why, using eight school data for a simple univariate case \citep{rubin1981} and twenty seven hospital profiling data for a bivariate case  \citep{everson2000inference}.  

\section{A Markov chain Monte Carlo method}\label{app1}

Our target distribution is the joint posterior distribution of the unknown parameters, $\boldsymbol{\theta}\equiv\{\boldsymbol{\theta}_1, \boldsymbol{\theta}_2, \ldots, \boldsymbol{\theta}_k\}$, $\boldsymbol{A}$, and $\boldsymbol{\beta}$, whose probability density function is defined as
\setcounter{equation}{4}\begin{equation}\label{joint_post}
\pi(\boldsymbol{\theta}, \boldsymbol{A}, \boldsymbol{\beta}\mid \boldsymbol{y})\propto h(\boldsymbol{A}, \boldsymbol{\beta})\times \prod_{j=1}^k\left[f(\boldsymbol{y}_j\mid \boldsymbol{\theta}_j)\times g(\boldsymbol{\theta}_j\mid \boldsymbol{A}, \boldsymbol{\beta})\right],
\end{equation}
where $\boldsymbol{y}=\{\boldsymbol{y}_1, \boldsymbol{y}_2, \ldots, \boldsymbol{y}_k\}$, the distributions for $f(\boldsymbol{y}_j\mid \boldsymbol{\theta}_j)$ and $g(\boldsymbol{\theta}_j\mid \boldsymbol{A}, \boldsymbol{\beta})$ are specified in \eqref{twolevel_structure1} and \eqref{twolevel_structure2}, respectively, and the joint prior density function $h(\boldsymbol{A}, \boldsymbol{\beta})$ is proportional to $\vert \boldsymbol{V}\!_0 + \boldsymbol{A}\vert ^{-p-1}\times I_{\{\vert \boldsymbol{A}\vert >0\}}d\boldsymbol{A}d\boldsymbol{\beta},$
meaning that the prior distribution on $\boldsymbol{A}$ is the USP and independently the prior distribution on $\boldsymbol{\beta}$ is an improper flat distribution (Lebesque) on $mp$-dimensional space. The joint posterior density in \eqref{joint_post}  is proper when $k>p+m+1$; see Theorem~2 of  \cite{everson2000inference}.

Since a rejection sampling for a multivariate Normal-Normal model  \citep{everson2000inference} is not available for various choices for the shape parameter of the USP, we use a Metropolis-Hasting within Gibbs sampler \citep{tierney1994markov}. This  sampler  iteratively samples the following conditional posterior distributions of the unknown parameters derived from \eqref{joint_post}; for  $j=1, 2, \ldots, k,$
\begin{align}
 \boldsymbol{\theta}_j\mid \boldsymbol{A}, \boldsymbol{\beta}, \boldsymbol{y}&\sim \textrm{N}_p\left(~(I_p-\boldsymbol{B}_j)\boldsymbol{y}_j+\boldsymbol{B}_j\boldsymbol{X}\!_j^\top \boldsymbol{\beta},~(I_p-\boldsymbol{B}_j)\boldsymbol{V}\!_j~\right),\label{conditional_theta}\\
\boldsymbol{\beta}\mid \boldsymbol{A},  \boldsymbol{\theta}, \boldsymbol{y}&\sim \textrm{N}_{mp}\left(\boldsymbol{\mu}_{\boldsymbol{\beta}},~\boldsymbol{\Sigma}_{\boldsymbol{\beta}}\right),\label{conditional_beta}\\
\pi_c(\boldsymbol{A}\mid \boldsymbol{\beta},  \boldsymbol{\theta}, \boldsymbol{y})&\propto \left[\prod_{j=1}^k\textrm{N}_p\left( \boldsymbol{\theta}_j\mid \boldsymbol{X}\!_j^\top \boldsymbol{\beta}, \boldsymbol{A}\right) \right]\times\vert \boldsymbol{V}\!_0 + \boldsymbol{A}\vert ^{-p-1}\times I_{\{\vert \boldsymbol{A}\vert >0\}},\label{conditional_A}
\end{align}
where $\boldsymbol{\mu}_{\boldsymbol{\beta}}=(\sum_{j=1}^k\boldsymbol{X}\!_j^\top \boldsymbol{A}^{-1}\boldsymbol{X}\!_j)^{-1}(\sum_{j=1}^k\boldsymbol{X}\!_j^\top \boldsymbol{A}^{-1}\boldsymbol{\theta}\!_j)$, $\boldsymbol{\Sigma}_{\boldsymbol{\beta}}=(\sum_{j=1}^k\boldsymbol{X}\!_j^\top \boldsymbol{A}^{-1}\boldsymbol{X}\!_j)^{-1}$, and the notation N$_p(w\mid \mu, \Sigma)$ used in \eqref{conditional_A} is a $p$-dimensional Gaussian density of $w$ with mean $\mu$ and covariance matrix $\Sigma$. At each iteration, we draw the unknown parameters in a sequence of \eqref{conditional_theta}, \eqref{conditional_beta}, and \eqref{conditional_A}. It is easy to sample the multivariate Gaussian conditional posterior distributions in \eqref{conditional_theta} and \eqref{conditional_beta}. However, the conditional posterior distribution of $\boldsymbol{A}$ in \eqref{conditional_A} is not a standard family distribution that we can directly sample, and thus we use a Metropolis-Hastings algorithm to sample \eqref{conditional_A} within the Gibbs sampler. (If we used an improper flat prior distribution on $\boldsymbol{A}$, the conditional posterior distribution of $\boldsymbol{A}$ in \eqref{conditional_A} would become an inverse Wishart$(k-p-1, \boldsymbol{S})$ distribution  with   degrees of freedom $k-p-1$ and scale matrix $\boldsymbol{S}\equiv\sum_{j=1}^k (\boldsymbol{\theta}_j-\boldsymbol{X}_j^\top\boldsymbol{\beta})(\boldsymbol{\theta}_j-\boldsymbol{X}_j^\top\boldsymbol{\beta})^\top$.) 

In a univariate case ($p=1$), we draw $\log(A^\ast)$ from a Gaussian proposal distribution N$_1(\log(A^{(i-1)})\mid \sigma^2)$ at iteration $i$, where the proposal scale $\sigma$ is the tuning parameter. We set $A^{(i)}$ to $A^\ast$ with probability 
\begin{equation}\label{accept_rate_uni}
\min\left[1,~\frac{\pi_c(A^\ast\mid \boldsymbol{\beta}^{(i)},  \boldsymbol{\theta}^{(i)}, \boldsymbol{y})}{\pi_c(A^{(i-1)}\mid \boldsymbol{\beta}^{(i)},  \boldsymbol{\theta}^{(i)}, \boldsymbol{y})}\times \frac{A^\ast}{A^{(i-1)}}\right]
\end{equation}
and set $A^{(i)}$ to $A^{(i-1)}$ otherwise. The last term in \eqref{accept_rate_uni}, $A^\ast/A^{(i-1)}$, is the Hastings ratio for the update of $A$ on a logarithmic scale. In a multivariate case, we set a  proposal distribution of $\boldsymbol{A}$ to an inverse Wishart distribution to avoid a matrix logarithm.  At iteration $i$, we draw $\boldsymbol{A}^\ast$ from the inverse Wishart($\nu$, $(\nu+p+1)\times\boldsymbol{A}^{(i-1)}$) distribution with degrees of freedom $\nu$ and scale matrix $(\nu+p+1)\times\boldsymbol{A}^{(i-1)}$; the mode of this inverse Wishart distribution is $\boldsymbol{A}^{(i-1)}$. We set $\boldsymbol{A}^{(i)}$ to $\boldsymbol{A}^\ast$  with probability
\begin{equation}
\min\left[1,~\frac{\pi_c(\boldsymbol{A}^\ast\mid \boldsymbol{\beta}^{(i)},  \boldsymbol{\theta}^{(i)}, \boldsymbol{y})}{\pi_c(\boldsymbol{A}^{(i-1)}\mid \boldsymbol{\beta}^{(i)},  \boldsymbol{\theta}^{(i)}, \boldsymbol{y})}\times \frac{\textrm{inv-W}(\boldsymbol{A}^{(i-1)}\mid \nu,~ (\nu+p+1)\times\boldsymbol{A}^\ast)}{\textrm{inv-W}(\boldsymbol{A}^\ast\mid \nu,~ (\nu+p+1)\times\boldsymbol{A}^{(i-1)})}\right],\nonumber
\end{equation}
where the notation $\textrm{inv-W}(\boldsymbol{A} \mid \nu,~ (\nu+p+1)\times \Sigma)$ denotes the density function of the inverse Wishart($\nu$, $(\nu+p+1)\times\Sigma$) distribution evaluated at $\boldsymbol{A}$. We treat the degrees of freedom $\nu$ as a tuning parameter for a reasonable acceptance rate. 

In our numerical illustrations, we run the Metropolis-Hastings within Gibbs sampler for 42000 iterations, discarding the first 2000 iterations as burn-in. We then choose every  other iteration of the Markov chain of length 40000, resulting in 20000 posterior samples of each parameter.

\section{Multivariate frequency coverage evaluation}\label{sec3}
To check which choice for the shape parameter of the USP produces Bayesian interval estimates for random effects that meet their nominal confidence levels better, we conduct a repeated sampling coverage evaluation. Here we extend a univariate repeated sampling coverage evaluation specified in \cite{tak2016b} to a multivariate one.  This evaluation is based on an assumption that our two-level Normal-Normal model is a true data generation process. We first generate $n_{\textrm{sim}}$ sets of $k$ random effects, $\{(\boldsymbol{\theta}^{(i)}_1, \boldsymbol{\theta}^{(i)}_2, \ldots, \boldsymbol{\theta}^{(i)}_k),~  i=1, 2, \ldots, n_{\textrm{sim}}\}$, from the conjugate Gaussian prior distribution in \eqref{twolevel_structure2} given specific values of $\boldsymbol{A}$ and $\boldsymbol{\beta}$. Let us denote these generative values of $\boldsymbol{A}$ and $\boldsymbol{\beta}$  by $\boldsymbol{A}_{\textrm{gen}}$ and $\boldsymbol{\beta}_{\textrm{gen}}$, respectively. Given these random effects, we generate $n_{\textrm{sim}}$ sets of $k$ unbiased estimates $\{(\boldsymbol{y}^{(i)}_1, \boldsymbol{y}^{(i)}_2, \ldots, \boldsymbol{y}^{(i)}_k),~  i=1, 2, \ldots, n_{\textrm{sim}}\}$  from the Gaussian distribution in \eqref{twolevel_structure1}.

We fit a Normal-Normal model on each simulation $i$, drawing 20000 posterior samples of $\boldsymbol{\theta}^{(i)}_1, \boldsymbol{\theta}^{(i)}_2, \ldots, \boldsymbol{\theta}^{(i)}_k, \boldsymbol{A}$, and $\boldsymbol{\beta}$ using the Metropolis-Hasting within Gibbs sampler.  For the $l$th component of $\boldsymbol{\theta}^{(i)}_j$, i.e., $\theta_{lj}^{(i)}$ for $l=1, 2, \ldots, p$, we calculate its 95\% posterior interval by 0.025 and 0.975 quantiles of its 20,000 posterior samples that we denote by $(\theta^{(i)}_{lj, \textrm{low}},~  \theta^{(i)}_{lj, \textrm{upp}})$. We define a coverage indicator of random effect~$j$ on the $i$-th simulation as 
\setcounter{equation}{9}\begin{equation}\label{coverage_indicator_vector}
I\left(\boldsymbol{\theta}_j^{(i)}\right)=\prod_{l=1}^p I\left(\theta_{lj}^{(i)}\right),
\end{equation} 
where $I(\theta_{lj}^{(i)})$ on the right hand side represents another indicator function that  takes the value one if the generative true value $\theta_{lj}^{(i)}$ lies  within its 95\% interval estimate $(\theta^{(i)}_{lj, \textrm{low}},~  \theta^{(i)}_{lj, \textrm{upp}})$ and zero otherwise. Thus the coverage indicator $I(\boldsymbol{\theta}_j^{(i)})$ in \eqref{coverage_indicator_vector} is equal to the value one if  all the $p$ components of the generative values $\boldsymbol{\theta}^{(i)}_j$ are within their 95\% posterior intervals and zero otherwise.

We assume that there exists an unknown coverage probability of random effect~$j$ denoted by $C_{\boldsymbol{A}_{\textrm{gen}},~ \boldsymbol{\beta}_{\textrm{gen}}}(\boldsymbol{\theta}_{j})$ depending on the values of $\boldsymbol{A}$ and  $\boldsymbol{\beta}$ that generate random effects and simulated data sets. The coverage indicator $I(\boldsymbol{\theta}_j^{(i)})$ in  \eqref{coverage_indicator_vector} is assumed to follow an independent and identically distributed  Bernoulli distribution given the unknown coverage rate $C_{\boldsymbol{A}_{\textrm{gen}},~ \boldsymbol{\beta}_{\textrm{gen}}}(\boldsymbol{\theta}_{j})$. The mean of the coverage indicators averaged over $n$ simulations, i.e., $\bar{I}(\boldsymbol{\theta}_{j})=\sum_{i=1}^{n_{\textrm{sim}}} I(\boldsymbol{\theta}_{j}^{(i)})/ n_{\textrm{sim}}$,  is a simple unbiased coverage estimator for $C_{\boldsymbol{A}_{\textrm{gen}},~ \boldsymbol{\beta}_{\textrm{gen}}}(\boldsymbol{\theta}_{j})$.  We define a more efficient Rao-Blackwellized (RB) unbiased coverage estimator as 
\begin{align}
\begin{aligned}
\bar{I}^{RB}(\boldsymbol{\theta}_{j}) &= \frac{1}{n_{\textrm{sim}}}\sum_{i=1}^{n_{\textrm{sim}}}E\!\left(I(\boldsymbol{\theta}_{j}^{(i)})~\big\vert~ \boldsymbol{A}_{\textrm{gen}}, \boldsymbol{\beta}_{\textrm{gen}}, \boldsymbol{y}_j^{(i)}\right)\\
&= \frac{1}{n_{\textrm{sim}}}\sum_{i=1}^{n_{\textrm{sim}}} P\bigg(\theta_{1j}^{(i)}\in\left(\theta_{1j, \textrm{low}}^{(i)}, \theta_{1j, \textrm{upp}}^{(i)}\right), \ldots, \\
&~~~~~~~~~~~~~~~~~~~~\theta_{pj}^{(i)}\in\left(\theta_{pj, \textrm{low}}^{(i)}, \theta_{pj, \textrm{upp}}^{(i)}\right)~\bigg\vert~ \boldsymbol{A}_{\textrm{gen}}, \boldsymbol{\beta}_{\textrm{gen}}, \boldsymbol{y}_j^{(i)}\bigg).
\end{aligned}\label{RB2}
\end{align}
The expectation of $\bar{I}^{RB}(\boldsymbol{\theta}_{j})$ given $\boldsymbol{A}_{\textrm{gen}}$ and $\boldsymbol{\beta}_{\textrm{gen}}$ is $C_{\boldsymbol{A}_{\textrm{gen}},~ \boldsymbol{\beta}_{\textrm{gen}}}(\boldsymbol{\theta}_{j})$. We use  the cumulative distribution of the  conditional posterior distribution of  random effect~$j$ in  \eqref{multivariate_conditional_posterior} to compute the conditional posterior probabilities in \eqref{RB2}. The variance of  $\bar{I}^{RB}(\boldsymbol{\theta}_{j})$ does not exceed that of $\bar{I}(\boldsymbol{\theta}_{j})$ because of the Rao-Blackwellization \citep{radhakrishna1945information, blackwell1947conditional}, and its unbiased estimate is 
\begin{equation}
\widehat{\textrm{Var}}\left(\bar{I}^{RB}(\boldsymbol{\theta}_{j})\right)=\frac{1}{n_{\textrm{sim}}(n_{\textrm{sim}}-1)}\sum_{i=1}^{n_{\textrm{sim}}}\left(E\!\left(I(\boldsymbol{\theta}_{j}^{(i)})~\big\vert~ \boldsymbol{A}_{\textrm{gen}}, \boldsymbol{\beta}_{\textrm{gen}}, \boldsymbol{y}_j^{(i)}\right)-\bar{I}^{RB}(\boldsymbol{\theta}_{j}) \right)^2\!.\nonumber
\end{equation}
To summarize the results of the frequency coverage evaluation, we report the following overall   coverage estimate and its variance estimate:
\begin{equation}\label{overall_RB}
\bar{\bar{I}}^{RB} = \frac{1}{k}\sum_{j=1}^k\bar{I}^{RB}(\boldsymbol{\theta}_{j})~~\textrm{and}~~\widehat{\textrm{Var}}(\bar{\bar{I}}^{RB})=\frac{1}{k^2}\sum_{j=1}^k\widehat{\textrm{Var}}\left(\bar{I}^{RB}(\boldsymbol{\theta}_{j})\right).
\end{equation}


\section{Numerical illustrations}\label{sec4}
Using the frequency coverage evaluation, we show which choice for the shape parameter produces Bayesian interval estimates for random effects that meet their nominal confidence levels better. The first numerical illustration is to investigate the shape parameter $V_0$ of the univariate USP ($p=1$) in a simple setting and the second one is to see whether this univariate case can be consistently extended to a multivariate USP ($p=2$).

\subsection{Eight school data}\label{sec4school}
To test the effect of coaching on students' SAT scores, the Education Testing Service  conducted randomized experiments in eight  schools \citep{rubin1981}. The data are composed of the estimated coaching effects on SAT scores ($y_{j}$) and standard errors  of the eight schools ($j=1, 2, \ldots, 8$). The data appear in Table~\ref{table1}.  We assume each school's coaching effect approximately follows a Gaussian sampling distribution and its sampling variance ($V_j$) is approximately known, considering a large number of students tested in each school \citep{tak2016b}. Since no covariate information is available, we assume $\boldsymbol{X}\!_j^\top \boldsymbol{\beta}=\beta_1$,  leading to an exchangeable Gaussian prior distribution in~\eqref{twolevel_structure2}. 

\begin{table}[t!]
\begin{center}\caption{The data are composed of the effect of coaching on students' SAT scores ($y_j$) and its standard error ($V_{j}^{1/2}$) for each of the eight schools ($j=1, 2, \ldots, 8$). }\label{table1}
\begin{tabular}{ccccccccc}
$j$ &1&2&3&4&5&6&7&8\\
\hline
$y_j$& 28 & 8 & $-3$ & 7 & $-1$ & 1 & 18 & 12\\
$V_j^{1/2}$& 15 & 10 & 16 & 11 & 9 & 11 & 10 & 18\\
\end{tabular}\end{center}
\end{table}

We begin the  frequency coverage evaluation by choosing the generative values, $A_{\textrm{gen}}$ and $\beta_{1, \textrm{gen}}$, in a way to favor the existent choice for $V_0$, e.g., the harmonic mean of $\{V_1, V_2, \ldots, V_k\}$, denoted by $V_{0, \textrm{DM}}$  \citep{dumouchel1994hierarchical}.  To see the pattern of coverage rates according to $A_{\textrm{gen}}$, we choose ten values of $A_{\textrm{gen}}$  by converting ten values of the shrinkage factor $B_0\in\{0.05, 0.15, \ldots, 0.95\}$, using the relationship $A=(1-B_0)V_{0}/B_0$. The resulting ten values of $A_{\textrm{gen}}$ with $V_{0}$ set to $V_{0, \textrm{DM}}=132.6$  are \{2520.2, 751.7, 397.9, 246.3, 162.1, 108.5, 71.4, 44.2, 23.4, 7.0\} and we denote these by $A_{\textrm{gen}, 1}$, $A_{\textrm{gen}, 2}$, $\ldots$, $A_{\textrm{gen}, 10}$. To set $\beta_{1, \textrm{gen}}$, we fit a Normal-Normal model using a USP with $V_{0, \textrm{DM}}$, and  set $\beta_{1, \textrm{gen}}$ to 7.95, the posterior mean of $\beta_1$ based on its 100,000 posterior samples.  Finally we have  ten pairs of the generative values, i.e., $(A_{\textrm{gen}, 1},  \beta_{1, \textrm{gen}})$, $(A_{\textrm{gen}, 2},  \beta_{1, \textrm{gen}})$, $\ldots$, $(A_{\textrm{gen}, 10},  \beta_{1, \textrm{gen}})$.

We compare the frequency coverage properties of six different prior distributions on~$A$;  the improper flat prior distribution on $A$ and the USPs on $A$ with five different shape parameters, i.e., $V_0= \delta \times V_{0, \textrm{DM}}$ for $\delta\in\{10^0, 10^1, 10^2, 10^3, 10^4\}$. For each prior distribution on $A$, we conduct ten frequency coverage evaluations using ten different pairs of the generative values, i.e., we conduct the frequency coverage evaluation 60 times in total for the combinations of ten pairs of the generative values and six different prior distributions on $A$. For each frequency coverage evaluation, we generate $1000~(=n_{\textrm{sim}})$ mock data sets. On each of the generated data sets, we fit a Normal-Normal model equipped with the corresponding prior distribution on $A$ and an improper flat prior distribution on $\beta_1$, drawing  20000 posterior samples of $\theta_1, \theta_2, \ldots, \theta_8, A$, and $\beta_1$. To initialize the Metropolis-Hastings within Gibbs sampler, we set $\theta_j^{(0)}=y_j$ for all~$j$, $\beta_1^{(0)}=\bar{y}$, and $A^{(0)}=V_{0, \textrm{DM}}$.  We set the scale of the Gaussian proposal distribution for $\log(A)$ to two, i.e., $\sigma=2$, resulting in an average acceptance rate equal to 0.326 across all the simulations. 

\begin{table}[t!]
\begin{center}\caption{The estimated coverage rates and their standard errors (in parentheses) specified  in \eqref{overall_RB}. The confidence level is set to 0.95. Three different prior distributions on $A$ appear in columns and ten different generative values of $B_{0, i}=V_{0, \textrm{DM}}/(V_{0, \textrm{DM}}+A_{\textrm{gen}, i})$ appear in rows. As the shape parameter of the USP increases, the coverage estimates based on the USP approach those based on the improper flat prior on $A$, conservatively achieving the 95\% confidence level.}\label{table1_5}
\begin{tabular}{lccc}
 & USP with  & USP with& \multirow{2}{*}{Improper flat on $A$}\\
 & $V_0=V_{0, \textrm{DM}}$ & $V_0=10^4 \times V_{0, \textrm{DM}}$ &  \\
\hline 
 $B_{0, 1}=0.05$ & 0.943 (0.0002) & 0.950 (0.0001) & 0.950 (0.0001) \\
 $B_{0, 2}=0.15$ & 0.931 (0.0005) & 0.951 (0.0002) & 0.950 (0.0002)\\
 $B_{0, 3}=0.25$ & 0.926 (0.0005)& 0.952 (0.0003)& 0.953 (0.0002) \\
 $B_{0, 4}=0.35$ & 0.930 (0.0005) & 0.955 (0.0003)& 0.954 (0.0003)\\
 $B_{0, 5}=0.45$ & 0.936 (0.0005) & 0.959 (0.0003)& 0.960 (0.0003)\\
 $B_{0, 6}=0.55$ & 0.944 (0.0004) & 0.963 (0.0004)& 0.965 (0.0004) \\
 $B_{0, 7}=0.65$ & 0.956 (0.0004) & 0.969 (0.0005)& 0.969 (0.0005)\\
 $B_{0, 8}=0.75$ & 0.967 (0.0005) & 0.974 (0.0007)& 0.974 (0.0006)\\
 $B_{0, 9}=0.85$ & 0.977 (0.0006) & 0.980 (0.0007)& 0.980 (0.0007)\\
 $B_{0, 10}=0.95$ &0.988 (0.0007) & 0.983 (0.0010)& 0.985 (0.0009)\\
\end{tabular}\end{center}
\end{table}

\begin{figure}[b!]
\begin{center}
\includegraphics[width = 380pt, height = 131pt]{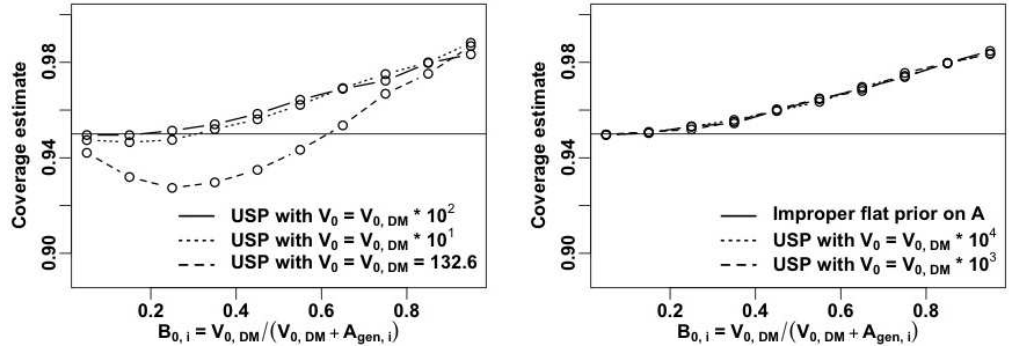}
\caption[]{Estimated coverage rates obtained from the frequency coverage evaluations. The overall coverage estimates based on $V_{0, \textrm{DM}}=132.6$, denoted by the dashed line on the left panel, do not meet the 95\% confidence level over a wide range of the generative values of $B_{0, i}$ between 0.05 and 0.55. As the USP on $A$ approaches the improper flat prior on $A$, i.e., as $V_0$ increases, the estimated coverage rates of the USP converge to those of the improper flat prior on $A$,  meeting the 95\% confidence level better.} 
\label{figure1}
\end{center}
\end{figure}

Using the 20000 posterior samples of each random effect, whose effective sample size is 8872 on average across all the simulations, we calculate 95\% posterior intervals for random effects, i.e., $(\theta^{(i)}_{j, \textrm{low}},~  \theta^{(i)}_{j, \textrm{upp}})$ for $j=1, 2, \ldots, 8$ and $i=1, 2, \ldots, 20000$. (Subscript $l$ used in \eqref{coverage_indicator_vector} is suppressed for simplicity because $p=1$.) Then, we calculate the RB coverage rate estimates, $\bar{I}^{RB}(\theta_j)$ in \eqref{RB2}. We summarize the overall  coverage  estimates  and their standard errors defined in \eqref{overall_RB} for three different prior distributions on $A$ in Table~\ref{table1_5} and display the overall coverage estimates for all 60 cases in Figure~\ref{figure1}. The standard errors  are  too small to be displayed.

In the left panel of Figure~\ref{figure1}, the overall coverage estimates based on the USP with the existent choice $V_{0, \textrm{DM}}$, denoted by the dashed line, do not achieve the 95\% confidence level over a wide range of the generative values of $B_{0, i}=V_{0, \textrm{DM}}/(V_{0, \textrm{DM}}+A_{\textrm{gen}, i})$ between 0.05 and 0.55. However, as the USP on $A$ approaches the improper flat prior on $A$, i.e., as $V_0$ increases, the estimated coverage rate of the USP converges to that of the improper flat prior on $A$, achieving the 95\% confidence level better; see the right panel of Figure~\ref{figure1}. This is because the USP, $(V_0 +A)^{-2}dA$, puts more mass at $A=0$ \emph{a priori} when $V_0$ is small.  Since the resulting posterior inference can be affected strongly by the prior distribution when the data size is small, the smaller shape parameter $V_0$ in the USP results in a smaller posterior sample of $A$, a larger posterior sample of  $B_j=V_j / (V_j+A)$, and thus  shorter posterior intervals for $\theta_j$ as implied by the variance in~\eqref{multivariate_conditional_posterior}. Consequently, these shorter posterior intervals produce smaller coverage estimates. The opposite occurs by flattening the USP on $A$ with larger $V_0$; the resulting posterior intervals for random effects become wider, leading to more conservative posterior inference on random effects.


The right panel of Figure~\ref{figure1} shows that the estimated coverage rates do not change much when $\delta$ in $V_0=\delta\times V_{0, \textrm{DM}}$ is greater than or equal to $10^3$. This indicates that a safe choice for $V_0$ to achieve good frequency coverage properties in the random effects estimation is to set a large enough value for $V_0$ compared to a typical value of $V_j$'s, e.g., $V_0 = 10^4 \times V_{0, \textrm{DM}}$ or $V_0 = 10^4 \times V_{0, \textrm{E\&M}}$, where $V_{0, \textrm{E\&M}}$ is the arithmetic mean of $\{V_1, V_2, \ldots, V_k\}$ suggested by \cite{everson2000inference}.

\begin{table}[b!]
\begin{center}\caption{The twenty seven hospital profiling data are composed of the  percentage of non-surgical problems ($Y_{1j}$) experienced by patients in hospital $j$,  that of surgical problems ($Y_{2j}$), the severity measure averaged over the health indices of interviewees ($x_{2j}$), and the number of interviewees ($n_j$). This table is reproduced from \cite{everson2000inference}. }\label{table2}
\begin{tabular}{ccccccccccc}
$j$ & $Y_{1j}$ & $Y_{2j}$ & $x_{2j}$  & $n_{j}$ & $~~~$& $j$ &$Y_{1j}$ & $Y_{2j}$ & $x_{2j}$  & $n_{j}$\\
\hline
1& 10.18 & 15.06 & 0.75 & 24 & & 15 & 15.80 & 11.50  & 0.26 & 61\\
2& 11.55 & 17.97 & 0.62 & 32 &  & 16 & 14.81 & 20.56 & 0.56 &62\\
3& 16.21 & 12.50 & 0.66 & 32 &  & 17 & 11.14 & 13.02 & 0.02 &62\\
4& 12.31 & 14.88 & 0.26 & 43 &  & 18 & 17.12 & 14.60 & 0.41&66\\
5& 12.88 & 15.21 & 0.96 & 44 &  & 19 & 16.93 & 16.28 & 0.56 &68\\
6& 11.84 & 17.69 & 0.44 & 45 &  & 20 & 11.02 & 13.52 & 0.34 &68\\
7& 14.82 & 16.91 & 0.44 & 48 &  & 21 & 14.69 & 16.49 & 0.56 &72\\
8& 13.05 & 15.07 & 0.55 & 49 &  & 22 & 10.48 & 14.24 & 0.79 &77\\
9& 12.43 & 12.01 & 0.33 & 51 &  & 23 & 15.82 & 15.13 & 0.47 &87\\
10& \multicolumn{1}{r}{8.35} & \multicolumn{1}{r}{9.43} & 0.47 & 53 &  & 24 & 12.66 & 14.99 & 0.71 & 122\\
11& 17.97 & 26.82 & 0.48 & 56 &  & 25 & 10.41 & 17.25 & 0.45 & 124\\
12& 11.84 & 15.64 & 0.34 & 58 &  & 26 & 10.32 & 10.13 & 0.05 & 149\\
13& 12.43 & 13.94 & 0.28 & 58 &  & 27 & 13.72 & 18.18 & 0.77 & 198\\
14& 14.73 & 15.40 & 0.63 & 60 &  &  & & & & \\
\end{tabular}\end{center}
\end{table}

\subsection{Twenty seven hospital profiling data}\label{hostpital}

The twenty seven hospital profiling data are based on the interviews of 1869 patients from  teaching hospitals and academic health centers \citep{cleary1991patients}. The data are composed of the number of interviewees ($n_j$), the severity measure averaged over the health indices of interviewees ($x_{2j}$), the  percentage of non-surgical problems ($Y_{1j}$), and that of surgical problems ($Y_{2j}$) experienced by patients in hospital $j$  ($j=1, 2, \ldots, 27$). (We fit an intercept term by setting $x_{1j}=1$ for all $j$.) The data are tabulated in Table~\ref{table2}.  \cite{everson2000inference} used these data to show the effectiveness of their rejection sampling based on the improper flat prior on $\boldsymbol{A}$   compared to the REML method. They estimated a common covariance matrix of the  percentages of non-surgical and surgical problems from the entire date of the 1869 patients and set it to $\boldsymbol{\Sigma}$ as follows;
\begin{equation}\nonumber
\boldsymbol{\Sigma} = 
\left( \begin{array}{cc}
148.87 & 140.43\\
140.43 & 490.60
\end{array} \right)\!.
\end{equation}
Using this $\boldsymbol{\Sigma}$, they set the covariance matrix of hospital $j$ to $\boldsymbol{V}\!_j= \boldsymbol{\Sigma}/n_j$. Following \cite{everson2000inference}, we  assume that the sampling distribution of ($Y_{1j}, Y_{2j}$) approximately follows a bivariate Gaussian distribution with known covariance matrix $\boldsymbol{V}\!_j$, considering a reasonably large number of interviewees in each hospital.

We start the  frequency coverage evaluation by setting $\boldsymbol{A}_{\textrm{gen}}$ and $\boldsymbol{\beta}_{\textrm{gen}}$ in such a way to favor the existent choice for $\boldsymbol{V}\!_0$, e.g., an arithmetic mean of $\{\boldsymbol{V}\!_1, \boldsymbol{V}\!_2, \ldots, \boldsymbol{V}\!_k\}$, denoted by $\boldsymbol{V}\!_{0, \textrm{E\&M}}$  \citep{everson2000inference}.  We first fit a bivariate Normal-Normal model using a USP with $\boldsymbol{V}\!_{0, \textrm{E\&M}}$ and a flat prior distribution on $\boldsymbol{\beta}$  and set $\boldsymbol{\beta}_{\textrm{gen}}$ to the posterior means of $\boldsymbol{\beta}$ based on its 100000 posterior samples. We set $\boldsymbol{A}_{\textrm{gen}, i}=\boldsymbol{\Sigma}/u_i$ for $i=1, 2, \ldots, 10$, where $u_i$ is the $i$th element of (0.29,  0.63, 1.00, 1.45, 2.04, 2.87, 4.16, 6.47, 11.82, 38.50) so that the resulting ten values of $\vert \boldsymbol{B}_{0, i}\vert\equiv\vert\boldsymbol{V}\!_{0, \textrm{E\&M}}(\boldsymbol{V}\!_{0, \textrm{E\&M}}+\boldsymbol{A}_{\textrm{gen}, i})^{-1}\vert$  are (0.05, 0.15, \ldots, 0.95).  Finally, we obtain ten different combinations of the generative values, $(\boldsymbol{A}_{\textrm{gen}, 1},  \boldsymbol{\beta}_{\textrm{gen}}),  \ldots, (\boldsymbol{A}_{\textrm{gen}, 10},  \boldsymbol{\beta}_{\textrm{gen}})$.

\begin{figure}[b!]
\begin{center}
\includegraphics[width = 380pt, height = 131pt]{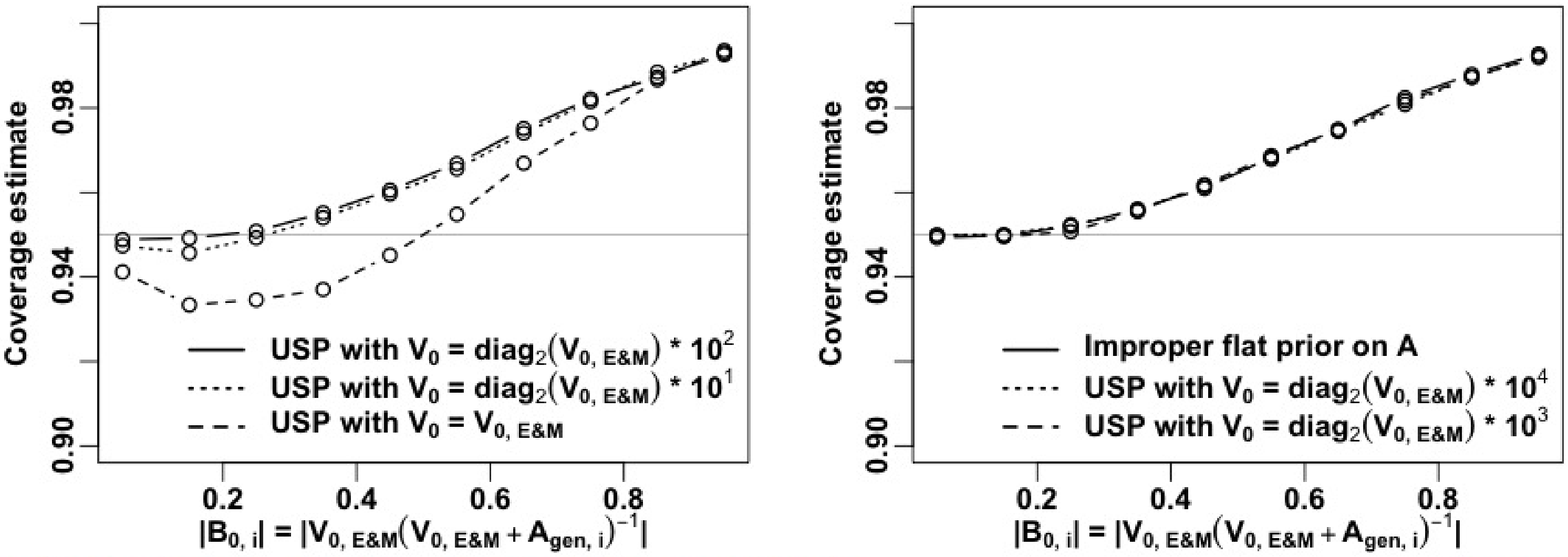}
\caption[]{Estimated coverage rates from the frequency coverage evaluation. The coverage estimates based on $V_{0, \textrm{E\&M}}$, denoted by the dashed line on the left panel exhibit under-coverages over a wide range of the generative values of $\vert\boldsymbol{B}_{0, i}\vert = \vert \boldsymbol{V}_{0, \textrm{E\&M}}(\boldsymbol{V}_{0, \textrm{E\&M}}+\boldsymbol{A}_{\textrm{gen}, i})^{-1}\vert$ between 0.05 and 0.55. As the USP on $\boldsymbol{A}$ approaches the improper flat prior distribution on $A$, i.e., as $\delta$ in $\boldsymbol{V}\!_0 = \delta\times \textrm{diag}_2(\boldsymbol{V}_{0, \textrm{E\&M}})$ increases, the estimated coverage rate of the USP converges to that of the improper flat prior distribution on $A$,  meeting the 95\% confidence level better. The notation $\textrm{diag}_p(\boldsymbol{V})$ represents a $p$-dimensional diagonal matrix whose diagonal elements are the same as those of $\boldsymbol{V}$.} 
\label{figure2}
\end{center}
\end{figure}

We compare operating characteristics of six different prior distributions on $\boldsymbol{A}$. We use  five USPs with five different shape parameters, i.e., $\boldsymbol{V}\!_0 =\boldsymbol{V}_{0, \textrm{E\&M}}$ and  $\boldsymbol{V}\!_0 = \delta\times \textrm{diag}_2(\boldsymbol{V}_{0, \textrm{E\&M}})$ for $w\in\{10^1, 10^2, 10^3, 10^4\}$, where the notation $\textrm{diag}_p(\boldsymbol{V})$ represents a $p$-dimensional diagonal matrix whose diagonal elements are the same as those of $\boldsymbol{V}$. The sixth prior distribution on $\boldsymbol{A}$ is an improper flat prior distribution, i.e., $I_{\{\vert \boldsymbol{A}\vert >0\}}d\boldsymbol{A}$. For each of the 60 frequency coverage evaluations (ten pairs of the generative values and six different prior distributions on $A$), we generate $1000~(=n_{\textrm{sim}})$ mock data sets. For each simulation, we fit a bivariate Normal-Normal model equipped with the corresponding prior distribution of $\boldsymbol{A}$ and $d\boldsymbol{\beta}$, drawing 20000 posterior samples of $\boldsymbol{\theta}$, $\boldsymbol{A}$, and $\boldsymbol{\beta}$ using the Metropolis-Hastings within Gibbs sampler. We set the initial value of $\boldsymbol{\theta}_j$ to $\boldsymbol{y}_j$ for all $j$,   of $\boldsymbol{\beta}$ to $\boldsymbol{\beta}_{\textrm{gen}}$, and of $\boldsymbol{A}$ to $\boldsymbol{V}_{0, \textrm{E\&M}}$. We set the degrees of  freedom $\nu$ of the inverse Wishart proposal distribution for $\boldsymbol{A}$ to 40, leading to the average acceptance rate equal to 0.378 across all the simulations. The average effective sample size for each random effect is 6817 across all the simulations. We calculate 95\% posterior intervals $(\theta^{(i)}_{lj, \textrm{low}},~  \theta^{(i)}_{lj, \textrm{upp}})$ for $l=1, 2$, $j=1, 2, \ldots, 27$ and $i=1, 2, \ldots, 20000$, and compute the RB coverage rate estimates, $\bar{I}^{RB}(\theta_j)$ in \eqref{RB2}. Table~\ref{table2_5} summarizes the overall  coverage estimates, $\bar{\bar{I}}^{RB}$ in  \eqref{overall_RB} and their standard errors for three different prior distributions. Figure~\ref{figure2} shows the overall coverage estimates for all 60 cases. The  standard errors are too small to be displayed.

\begin{table}[t!]
\begin{center}\caption{The  coverage estimates and their standard errors (in  parentheses) specified in~\eqref{overall_RB}. The confidence level is set to 0.95. Three different prior distributions on $A$ appear in columns and  ten  determinants of the generative values of $\boldsymbol{B}_{0, i}=\boldsymbol{V}_{0, E\&M}/(\boldsymbol{V}_{0, E\&M}+\boldsymbol{A}_{\textrm{gen}, i})$ appear in rows. As $\delta$ in $V_0=\delta\times\textrm{diag}_2(\boldsymbol{V}_{0, E\&M})$ increases, the coverage estimates based on the USP approach those based on the improper flat prior on $A$,  conservatively achieving the 95\% confidence level. The notation $\textrm{diag}_p(\boldsymbol{V})$ represents a $p$-dimensional diagonal matrix whose diagonal elements are the same as those of $\boldsymbol{V}$.}\label{table2_5}

\begin{tabular}{lccc}
 & USP with  & USP with& \multirow{2}{*}{Improper flat on $A$}\\
 & $\boldsymbol{V}_0=\boldsymbol{V}_{0, \textrm{E\&M}}$ & $\boldsymbol{V}_0=10^4 \times\textrm{diag}_2(\boldsymbol{V}_{0, \textrm{E\&M}})$ & \\
\hline 
 $\vert\boldsymbol{B}_{0, 1}\vert=0.05$ & 0.941 (0.0002) & 0.950 (0.0001) & 0.949 (0.0001) \\
 $\vert\boldsymbol{B}_{0, 2}\vert=0.15$ & 0.933 (0.0003) & 0.950 (0.0002) & 0.950 (0.0002) \\
 $\vert\boldsymbol{B}_{0, 3}\vert=0.25$ & 0.935 (0.0003) & 0.952 (0.0002)& 0.952 (0.0002)\\
 $\vert\boldsymbol{B}_{0, 4}\vert=0.35$ & 0.937 (0.0003) & 0.956 (0.0002)& 0.956 (0.0002)\\
 $\vert\boldsymbol{B}_{0, 5}\vert=0.45$ & 0.945 (0.0003) & 0.962 (0.0002)& 0.961 (0.0002)\\
 $\vert\boldsymbol{B}_{0, 6}\vert=0.55$ & 0.955 (0.0003) & 0.968 (0.0002)& 0.968 (0.0002)\\
 $\vert\boldsymbol{B}_{0, 7}\vert=0.65$ & 0.967 (0.0002) & 0.974 (0.0002)& 0.975 (0.0003)\\
 $\vert\boldsymbol{B}_{0, 8}\vert=0.75$ & 0.976 (0.0003) & 0.981 (0.0003)& 0.982 (0.0002)\\
 $\vert\boldsymbol{B}_{0, 9}\vert=0.85$ & 0.987 (0.0003) & 0.987 (0.0003)& 0.988 (0.0003)\\
 $\vert\boldsymbol{B}_{0, 10}\vert=0.95$ & 9.994 (0.0003)& 0.992 (0.0004)& 0.993 (0.0003)\\
\end{tabular}\end{center}
\end{table}

The results displayed in Figure~\ref{figure2} are similar to those of the univariate case ($p=1$) in Section \ref{sec4school}. In the left panel of Figure~\ref{figure2}, the coverage estimates based on the existent choice $\boldsymbol{V}_{0, \textrm{E\&M}}$, denoted by the dashed line, do not achieve the 95\% confidence level over the  range of the generative values of $\vert \boldsymbol{B}_{0, i}\vert= \vert \boldsymbol{V}_{0, \textrm{E\&M}}(\boldsymbol{V}_{0, \textrm{E\&M}}+\boldsymbol{A}_{\textrm{gen}, i})^{-1}\vert$ between 0.05 and 0.55. However, as the USP  approaches the improper flat prior on $\boldsymbol{A}$, i.e., as $\delta$ in $\boldsymbol{V}\!_0 = \delta\times \textrm{diag}_2(\boldsymbol{V}_{0, \textrm{E\&M}})$ increases, the estimated coverage rate of the USP approaches that of the improper flat prior on~$\boldsymbol{A}$ as shown in the right panel. This is because the large value of $\delta$ flattens the USP on~$\boldsymbol{A}$ and thus the coverage estimates based on the USP approach those based on the improper flat prior on~$\boldsymbol{A}$.


The right panel of Figure~\ref{figure2} shows that the estimated coverage rates do not change much when $\delta$ in $\boldsymbol{V}\!_0 = \delta\times \textrm{diag}_2(\boldsymbol{V}_{0, \textrm{E\&M}})$ is greater than or equal to $10^3$. This indicates that it is  safe to set a large enough value of $\delta$ that leads to a large value of $\vert \boldsymbol{V}\!_0\vert$ compared to a typical value of $\vert\boldsymbol{V}_j\vert$'s to achieve better frequency coverage properties in the random effects estimation, e.g.,  $\boldsymbol{V}\!_0 = 10^4\times \textrm{diag}_2(\boldsymbol{V}_{0, \textrm{E\&M}})$ or $\boldsymbol{V}_0 = 10^4\times \textrm{diag}_2(\boldsymbol{V}_{0, \textrm{DM}})$.


\section{Concluding Remarks}\label{sec5}
A uniform shrinkage prior (USP) on the unknown variance component of a random-effects model is known to produce good frequency properties \citep{christiansen1997hierarchical, daniels1999prior, natarajan2000reference}. However, it has been neglected whether the USP can maintain such good frequency properties regardless of its shape parameter. We implement frequency coverage evaluations to see which choice for the shape parameter of the USP enables the posterior intervals for random effects to meet their nominal confidence levels better. In our numerical illustrations,  the shape parameter that flattens the USP on $\boldsymbol{A}$ achieves better frequency coverage properties in estimating random effects than the existent choices suggested in the literature. 

There are several opportunities to build upon our work. Implementing the frequency coverage evaluation is computationally expensive because it fits a model on every simulated data set. For example,  it takes almost a week to complete one frequency coverage evaluation that fits the model on a thousand simulated data sets in  Section~\ref{hostpital}. Thus, it is desirable to develop an independent sampler for the Normal-Normal model with the USP instead of a Markov chain Monte Carlo method used in this article. Also, the frequency coverage properties for the USP according to the shape parameter have not been tested in different hierarchical models, e.g., Poisson-Gamma or Beta-Binomial models, though they may produce similar results  asymptotically. Another avenue for further improvement is to check high-dimensional behavior of the USP according to the shape parameter. We leave these for  our future research. 





\bibliographystyle{ba}
\bibliography{sample}

\begin{acknowledgement}
The author acknowledges partial support from the NSF under Grant DMS 1127914 to the Statistical and Applied Mathematical Sciences Institute and thanks Carl N. Morris for very helpful discussions and Steven R. Finch for proofreading.. 
\end{acknowledgement}

\end{document}